\documentclass[prl,twocolumn,showpacs,floatfix]{revtex4}

\usepackage{bm}
\usepackage{graphicx}  
\usepackage{latexsym}                
\usepackage{amsmath}     
\usepackage{amsfonts}  
\usepackage{amssymb}
\usepackage{amsthm}
\usepackage{dcolumn}
\usepackage{color}

\renewcommand{\phi}{\varphi}
\renewcommand{\>}{\right \rangle}

\newcommand{\ket}[1]{\left |#1\>}

\newcommand{\be}{\begin{equation}}
\newcommand{\ee}{\end{equation}}
\newcommand{\bea}{\begin{eqnarray}}
\newcommand{\eea}{\end{eqnarray}}

\begin{document}
\title{Directional Coupling for Quantum Computing and Communication}

\author{Georgios M. Nikolopoulos}
\affiliation{Institute of Electronic Structure and Laser, FORTH, P. O. Box 1527, Heraklion 711 10, Crete, Greece}

\date{\today}

\begin{abstract}
We introduce the concept of directional coupling, i.e., the selective 
transfer of a state between adjacent quantum wires, in the context of 
quantum computing and short-distance communication. Our analysis rests upon a 
mathematical analogy between a dual-channel directional coupler and 
a composite spin system.  

\end{abstract}


\maketitle

{\em Introduction.}--- 
Directional coupling, that is the exchange of power between guided modes 
of adjacent waveguides, has many applications in (opto)electronics 
\cite{book1,electronDC,pbg}. 
For instance, directional couplers (DCs) may perform a number of useful 
functions in thin-film devices such as, power division, switching, frequency selection, 
and (de)multiplexing. Typically, a dual-channel DC is a passive device with 
two input and two output ports. The ports are the ends of two 
waveguides, the so-called source 
and drain channels, which are brought in close proximity over a certain region. 
Varying a control parameter, one may achieve any division of a signal entering the source 
channel, between the outputs of the two channels.  

From the theoretical point of view, directional coupling can be treated in the framework of 
coupled-mode theory, where one deals with equations of motion for the complex amplitudes 
pertaining to the two forward-propagating guided modes \cite{book1,electronDC}. 
Usually, back reflection is absent 
due to technical reasons (e.g., in electronic devices an externally applied voltage may  
determine the propagation direction for electrons), while in certain configurations 
pertaining to optical waveguides, it has been shown that excitation of 
backward-propagating modes can be suppressed by applying adiabatic mode-coupling 
techniques \cite{back-reflection,STIRAP}, which require an elaborate sequence of pulses. 
 
In this Letter, we propose a dual-channel DC for quantum computing and short-distance communication purposes. 
In analogy to conventional DCs, a dual-channel quantum DC (QDC) can be defined as a 
device which allows the selective transfer of quantum signals (i.e., quantum states  
of information carriers \cite{Vinc00}), between two adjacent quantum channels. 
A QDC is expected to be the key element  
for various useful quantum information processing tasks, such as quantum switching, 
(de)multiplexing, etc.

The information carriers involved in some of the most promising 
proposals for large-scale quantum computing are not compatible with photons    
\footnote{For photonic carriers one may use optical fibers and DCs.}. 
Hence, the engineering of perfect quantum channels for specific information carriers  
has recently attracted considerable interest \cite{review}. 
In contrast to their (opto)electronic counterparts, quantum channels are 
discrete, as they typically 
pertain to arrays of coupled quantum objects (sites). An excitation created 
somewhere in the array, will unavoidably propagate 
in both directions and,  after some time, various sites of the array may be occupied 
with different probabilities. 
Despite such delocalization effects, it has been shown that an excitation 
can be transferred in a perfect and deterministic way between the two ends of the channel, 
by engineering the couplings between adjacent sites \cite{NPL04,CDEKL04}.  

Given two perfect quantum channels, our task here is to define 
inter-channel interactions, for which the entire system operates as a dual-channel QDC. 
In other words, we will discuss the conditions   
under which a ``flying'' qubit \cite{Vinc00}, prepared initially in the first site of 
one of the channels,   
can be transferred to the last site of either of the two channels, 
in a controlled and deterministic manner. 
We are interested in symmetric configurations with minimal external control, i.e., without 
elaborate sequences of time-dependent pulses and measurements.   


{\em A perfect quantum channel.} --- 
A Hamiltonian for perfect state-transfer (PST)
along a chain of $N$ coupled nearly identical sites is of the form $(\hbar =1)$
\begin{subequations}
\label{PST_ham}
\bea
\hat{{\cal H}}_{\rm PST} = \sum_{j=1}^{N} \varepsilon \hat{c}^\dag_{j} \hat{c}_{j} 
+\sum_{j=1}^{N-1}\Omega_{j,j+1}
(\hat{c}^\dag_{j} \hat{c}_{j+1}+\hat{c}^\dag_{j+1} \hat{c}_{j}),
\label{PST_ham1}
\eea
where, $\hat{c}^\dag_{j}$ is the creation operator for an excitation on the  
$j$th site of the channel with energy $\varepsilon$ 
\footnote{Typically, the excitation pertains to an information carrier 
prepared in a quantum state. This state may involve various 
degrees of freedom that we neglect here, 
assuming their preservation throughout the evolution of the system.}, 
and $\Omega_{j,j+1}$ is the coupling between adjacent sites. 
Let us consider a situation when a single excitation is prepared initially 
in the first site of the chain. The Hamiltonian (\ref{PST_ham1}) preserves the number of excitations  
and thus, the system is restricted to the one-excitation Hilbert space 
throughout its evolution.  Hence, the computational basis can 
be chosen as $\{\ket{j}\}$, where $\ket{j}\equiv\hat{c}^\dag_{j}\ket{\{0\}}$ 
is the state with one excitation on the $j$th site, 
and $\ket{\{0\}}$ denotes the vacuum state of the system. 
As has been shown in \cite{NPL04,CDEKL04}, the chain acts as a perfect quantum channel, 
i.e., one can achieve PST from the first to the last site, 
by judiciously engineering the coupling strengths according to 
\be
\Omega_{j,j+1}=\Omega \sqrt{(N-j)j}. 
\label{Omeg}
\ee
\end{subequations}
Moreover, setting $J=(N-1)/2$ and $m=j-(N+1)/2$, one may define a one-to-one 
correspondence between  the angular momentum (AM) basis $\{\ket{J,m}\}$ and 
the computational basis $\{\ket{j}\}$. In view of this correspondence, the evolution of 
the excitation under the influence of the PST Hamiltonian (\ref{PST_ham}), is analogous to the 
evolution of the spin-$J$ system,  which is rotated around the $x$-axis \cite{shore}. 
In the following we discuss how such a quantum channel can be used as a  building block for 
a QDC. Our analysis rests upon the aforementioned mathematical analogy, 
which turns out to be a rather useful theoretical tool.


{\em Quantum directional coupler.}--- 
In analogy to its (opto)electronic counterparts, a dual-channel QDC involves two nearly 
identical channels, the source (s) and the drain (d). 
Each channel consists of $N>2$ nearly identical sites 
denoted by $(\sigma,j)$, with $\sigma\in\{{\rm s},{\rm d}\}$  and $1\leq j\leq N$.  
Accordingly, the computational basis of the system can be chosen as 
$\{\ket{\sigma;j}\}$, with $\ket{\sigma;j}$ denoting 
an excitation on the $j$th site of channel $\sigma$. 
The first two sites $\{({\rm s},1); ({\rm d},1)\}$ play the role of the two input ports, 
whereas the output ports are represented by the last sites 
$\{({\rm s},N); ({\rm d},N)\}$. 

The source and the drain channels are described by a PST Hamiltonian of the 
form (\ref{PST_ham}), and our task is to define  
interactions between them so that an excitation initially occupying one of the 
input ports, can be transferred to either of the two output ports in a controlled and 
deterministic way. 
More precisely, consider the excitation initially occupying the first site of the 
source channel \footnote{For symmetric QDCs, it is sufficient to solve the problem with either of 
the two input ports initialized. 
}, i.e., the device is initially prepared in the state 
\be
\ket{\Psi_{\rm C}(0)}=\ket{{\rm s};1}.
\label{init_C}
\ee
At well defined time instants, a dual-channel QDC should be capable of performing perfectly 
the transformations
\begin{subequations}
\label{transC}
\bea
\label{transC1}
&&\ket{{\rm s};1}\to \ket{{\rm s};N}, \\
&&\ket{{\rm s};1}\to \ket{{\rm d};N}
\label{transC2}
\eea
\end{subequations}
apart, perhaps, from an unimportant global phase. 
Moreover, one should be able to switch between the two transformations, by adjusting a set 
of parameters controlling the inter-channel interactions. 

We discuss two different symmetric configurations of sites that may operate 
as a dual-channel QDCs. 
Both of them pertain to a grid of $M\times N$ nearly 
identical sites with pre-engineered couplings, although their operation relies 
on fundamentally different principles. In second quantization, 
the dynamics of a single excitation in such a two-dimensional structure is 
described by a Hamiltonian of the form  
\bea
\hat{\cal H}_{M\times N}=
\frac{1}{2}\sum_{i,i^\prime=1}^{M}~\sum_{j,j^\prime=1}^{N} G_{j,j^\prime}^{i,i^\prime} 
(\hat{a}^\dag_{i,j} \hat{a}_{i^\prime,j^\prime}+
\hat{a}^\dag_{i^\prime,j^\prime} \hat{a}_{i,j}), 
\label{Ham_grid}
\eea
where $\hat{a}^\dag_{i,j}$ creates an excitation on the $j$th site of the 
$i$th row with energy $\varepsilon=G_{j,j}^{i,i}$, while the coupling strength 
between two different sites $(i,j)$ and $(i^\prime, j^\prime)$,  
is denoted by  $G_{j,j^\prime}^{i,i^\prime}$, with $G_{j,j^\prime}^{i,i^\prime}=G_{j^\prime,j}^{i^\prime,i}$. 
In this formalism, the two outer most chains represent the source and the drain channels 
(i.e., ${\rm s}\equiv 1$ and ${\rm d}\equiv M\geq 2$), 
while any intermediate sites $(i,j)$ with $i\neq \{1,M\}$ pertain to the coupler. 
Depending on the particular quantum-computing realization under consideration, 
each site of the grid may correspond, for instance, to a quantum dot or a superconducting qubit.

In general, the two channels of a QDC may be coupled directly, or indirectly 
through their interaction with another system (coupler) placed between them. 
To describe the operation of the device in a unified theoretical framework, 
we may introduce two AM operators $\hat{\bf J}_{\rm h}$ and $\hat{\bf J}_{\rm v}$ 
acting on different subspaces, with 
\begin{subequations}
\label{cores}
\bea
\label{cores1}
&&J_{\rm h}=(N-1)/2,\quad m_{\rm h}=j-(N+1)/2;\\
&&J_{\rm v}=(M-1)/2,\quad m_{\rm v}=i-(M+1)/2.  
\label{cores2}
\eea
\end{subequations} 
An orthonormal basis for the state space of the spin-$J_{\rm \alpha}$ system 
(with $\alpha\in\{{\rm h,v}\}$), can be chosen as $\{\ket{J_\alpha,m_{\alpha}}\}$, 
where $\ket{J_\alpha,m_{\alpha}}$ are degenerate eigenvectors of the 
operator $\hat{J}^2_{\alpha}$. 
As we will see later on, this degeneracy specifies the class 
of QDCs, whose operation can be simulated by the dynamics of 
the composite spin system with basis states 
$\{\ket{J_{\rm v},m_{\rm v};J_{\rm h},m_{\rm h}}\}$,  
where $\ket{J_{\rm v},m_{\rm v};J_{\rm h},m_{\rm h}}\equiv 
\ket{J_{\rm v},m_{\rm v}}\otimes \ket{J_{\rm h},m_{\rm h}}$.

The role of the spin-$J_{\rm h}$ system is to describe the dynamics of 
the excitation in either of the two nearly identical channels (source or drain). 
For fixed channel parameters $\{N,\varepsilon,\Omega\}$, one may define a 
one-to-one correspondence between the basis states 
$\{\ket{j}\}$ and $\{\ket{J_{\rm h},m_{\rm h}}\}$, by means of Eqs. (\ref{cores1}), 
i.e.,  we have 
\begin{subequations}
\label{conv}
\be
\label{conv1}
\ket{J_{\rm h},m_{\rm h}}\equiv\ket{j}.
\ee 
On the other hand, the spin-$J_{\rm v}$ system has been introduced for the 
description of the inter-channel dynamics, with the only convention being  
\be
\ket{J_{\rm v},-J_{\rm v}}\equiv\ket{\rm s},\quad  
\ket{J_{\rm v},J_{\rm v}}\equiv\ket{\rm d}.
\label{conv2}
\ee 
\end{subequations} 
In view of Eqs. (\ref{cores2}) and (\ref{conv2}), 
a dual-channel QDC with 
directly coupled channels is described by a spin-$1/2$ particle (i.e., for $M=2$), 
whereas the presence of a coupler between the two channels is represented by a 
spin-$J_{\rm v}$ particle with $J_{\rm v}>1/2$ (i.e., for $M>2$).  
In the latter case, the spin states $\ket{J_{\rm v},m_{\rm v}}$ with $m_{\rm v}\neq\pm J_{\rm v}$  
correspond to the coupler.

In AM representation, the dynamics of a single excitation in a dual-channel QDC 
can be described by a Hamiltonian of the form 
\be
\hat{\cal H}=\hat{\cal H}_{\rm h}+\hat{\cal H}_{\rm v}, 
\label{H_tot}
\ee 
where $\hat{\cal H}_{\alpha}\equiv\hat{{\cal H}}_{\alpha}^{(0)}+\hat{{\cal V}}_{\alpha}$ 
refers to the spin-$J_{\alpha}$ system only.  
The basis states $\{\ket{J_{\alpha},m_{\alpha)}}\}$ 
are degenerate eigenstates of the corresponding unperturbed Hamiltonian 
$\hat{\cal H}_{\alpha}^{(0)}\equiv \varepsilon_{\alpha} \hat{J}^2_{\alpha}$, 
while $\hat{{\cal V}}_{\alpha}$ is the coupling between various states 
$\{\ket{J_{\alpha},m_{\alpha}}\}$. 
Due to the degeneracy, conventions (\ref{conv}) imply that a dual-channel QDC 
can be described in the present theoretical framework if 
the coupler is on resonance with both channels and, for a given channel, all 
the states $\ket{\sigma;j}$ have the same energy.

The initial condition (\ref{init_C}) reads in AM representation 
\be
\ket{\Psi_{\rm AM}(0)}=\ket{J_{\rm v},-J_{\rm v};J_{\rm h},-J_{\rm h}},
\label{init_AM}
\ee  
whereas transformations (\ref{transC}) are in one-to-one correspondence with the 
following transformations
\begin{subequations}
\label{transAM}
\bea
\label{transAM1}
&&\ket{J_{\rm v},-J_{\rm v};J_{\rm h},-J_{\rm h}}\to \ket{J_{\rm v},-J_{\rm v};J_{\rm h},J_{\rm h}},\\
&&\ket{J_{\rm v},-J_{\rm v};J_{\rm h},-J_{\rm h}}\to \ket{J_{\rm v},J_{\rm v};J_{\rm h},J_{\rm h}}.  
\label{transAM2}
\eea 
\end{subequations}

Transformation (\ref{transAM1}) pertains to the evolution of the  spin-$J_{\rm h}$ system 
only, and is thus expected to be implementable by Hamiltonian (\ref{H_tot}) for 
$\hat{{\cal V}}_{\rm v}=0$ (no inter-channel coupling). 
Recall now that the dynamics of the spin-$J_{\rm h}$ system have to describe accurately 
the evolution of the excitation in either of the two (nearly identical) channels.  
Given that both channels are described by a PST Hamiltonian of the 
form (\ref{PST_ham}), with fixed parameters $\{N,\varepsilon,\Omega\}$, 
we can easily specify the form of the Hamiltonian 
$\hat{{\cal H}}_{\rm h}$, by expressing the PST 
Hamiltonian (\ref{PST_ham}) in the basis $\{\ket{J_{\rm h},m_{\rm h}}\}$.
Using correspondences (\ref{cores1}) we find
$\hat{{\cal H}}_{\rm h} =\varepsilon_{\rm h}\hat{J}^2_{\rm h}+2\Omega \hat{J}_{{\rm h},x}$,
where $\varepsilon_{\rm h}=\varepsilon/[{J_{\rm h}(J_{\rm h}+1)}]$, 
and $\hat{J}_{{\rm h},x}$ is the $x$-component of the vector $\hat{\bf J}_{{\rm h}}$.  
Under the influence of $\hat{{\cal H}}_{\rm h}$, the initial state of the isolated spin-$J_{{\rm h}}$ 
system undergoes a rotation around the $x$-axis, and the transformation (\ref{transAM1}) 
takes place at time $\tau=\pi/(2\Omega)$. 
Having specified the first part of Hamiltonian (\ref{H_tot}), we have to determine the 
inter-channel interaction $\hat{{\cal V}}_{\rm v}$, for which 
transformation (\ref{transAM2}) takes place at a well defined time instant. 

The transformation (\ref{transAM2}) involves a simultaneous rotation of the initial states 
of both spins. This leads us to introduce the total angular momentum 
$\hat{\bf J}=\hat{\bf J}_{\rm h}+\hat{\bf J}_{\rm v}$, with 
$|J_{\rm h}-J_{\rm v}|\leq J\leq J_{\rm h}+J_{\rm v}$ and $|m|\leq J$, while 
the corresponding basis states $\{\ket{J,m}\}$ can be expanded on the basis 
$\{\ket{J_{\rm v},m_{\rm v};J_{\rm h},m_{\rm h}}\}$ in the usual way \cite{Merz68}. 
In the basis $\{\ket{J,m}\}$, the initial condition (\ref{init_AM}) reads 
$\ket{\Psi_{\rm AM}(0)}=\ket{J,-J}$, while for the transformation (\ref{transAM2}) we have 
$\ket{J,-J}\to \ket{J,J}$.
In view of the previous discussion, this transformation can be performed by defining the 
inter-channel coupling $\hat{\cal V}_{\rm v}$, 
such that $\hat{\cal H}\sim 2\Omega\hat{J}_x$. One may choose 
$\hat{\cal V}_{\rm v}=2K \hat{J}_{{\rm v},x}$, with $K$ denoting the inter-channel coupling strength. 
The total Hamiltonian (\ref{H_tot}) then reads  
\be
\hat{\cal H}=\varepsilon_{\rm h}\hat{J}^2_{\rm h} +\varepsilon_{\rm v}\hat{J}^2_{\rm v}
+2\Omega \hat{J}_{{\rm h},x}+2K \hat{J}_{{\rm v},x}, 
\label{ham_DC}
\ee
and acquires the desired form for $K=\Omega$.  

The Hamiltonian (\ref{ham_DC}) describes the operation of a perfect dual-channel QDC in an 
AM representation. Hence, a quantum network involving 
a number of coupled sites,  may operate as a QDC if the Hamiltonian of the entire system 
in an AM representation acquires the form (\ref{ham_DC}). 
For instance, one can readily show, using Eqs. (\ref{cores}), that Hamiltonian (\ref{Ham_grid}) 
reduces to Hamiltonian (\ref{ham_DC}) when adjacent sites are coupled with strengths 
$G_{j,j+1}^{i,i}=\Omega\sqrt{j(N-j)}$ and $G_{j,j}^{i,i+1}=K\sqrt{i(M-i)}$; which incidentally 
underscores the usefulness of the AM representation. 
This coupling configuration has also been investigated in \cite{Li07}, 
albeit in a different context. The present work, however, reveals another aspect of such a 
structure, namely its use as a QDC with source and drain channels the two outer most chains, 
and control parameter $K$. In particular, a qubit state can be transferred selectively from the input port 
$({\rm s},1)$, to either of the two output ports $\{({\rm s},N),({\rm d},N)\}$ 
at time $t=\tau$ for $K=\{0,\Omega\}$, respectively. 

The main disadvantage of this configuration, however, is that all the sites of the source channel 
have to be coupled (directly or indirectly) to the corresponding sites of the drain channel 
via nearest-neighbor interactions. Depending on the physical realization under consideration, 
this might be very restrictive as it may imply that the two quantum channels have to be close 
to each other.  The question therefore is, can one achieve directional coupling between two 
chains by defining inter-channel interactions for a certain number of sites only?

This question cannot be answered in the framework of Hamiltonian (\ref{H_tot}), 
as it describes independent evolutions of the vectors $\hat {\bf J}_{\rm v}$ and 
$\hat {\bf J}_{\rm h}$. Instead, one has to consider more general Hamiltonians involving 
coupled angular momenta. Consider, for instance, an inter-channel 
interaction $\hat{\cal W}$, represented by the coupling between a spin $J_{{\rm v}}=1/2$ and an angular 
momentum $J_{{\rm h}}=1$,  
such that $\hat{\cal W}=K\hat{J}_{{\rm v},y}\hat{J}_{{\rm h},y}$. 
In view of the previous discussion, we may write the total Hamiltonian   
\be
\hat{\cal H}=\varepsilon_{\rm h}\hat{J}^2_{\rm h} +\varepsilon_{\rm v}\hat{J}^2_{\rm v}
+2\Omega \hat{J}_{{\rm h},x} +K\hat{J}_{{\rm v},y}\hat{J}_{{\rm h},y}, 
\label{ham_DC2}
\ee
and transformation (\ref{transAM1}) can be achieved for $K=0$, 
when only the vector $\hat{\bf J}_{{\rm h}}$ is rotated around the $x$-axis. 
Turning on the inter-channel interaction, i.e., setting $K\neq 0$, 
both vectors $\hat{\bf J}_{{\rm h}}$ and $\hat{\bf J}_{{\rm v}}$ can be rotated simultaneously 
around the $y$-axis. In this case, the initial state of the system  
evolves under the influence of both $\hat{\cal V}_{\rm h}$ and $\hat{\cal W}$. 
Hence,  we have two distinct evolution routes 
that may interfere either constructively or destructively, and the 
transformation (\ref{transAM2}) can be achieved by choosing judiciously the ratio $K/\Omega$. 
Indeed, for the initial condition (\ref{init_AM}), one can show that the transformation 
(\ref{transAM2}) occurs at $t=\tau/\sqrt{2}$ for $K=-4\Omega$. 

\begin{figure}[h]
\includegraphics[width=7.cm]{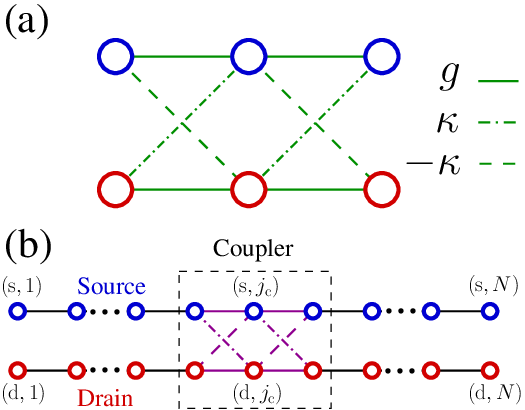}
\caption{
(color online). (a) A coupler consisting of six identical sites. 
(b) The coupler integrated in a dual-channel system. 
} \label{setup.fig}
\end{figure}

In general, Hamiltonian (\ref{ham_DC2}) can be implemented in the $2\times 3$ grid 
depicted in Fig. \ref{setup.fig}(a), for $g=\sqrt{2}\Omega$ and $\kappa=K/(2\sqrt{2})$. 
Such a configuration can be used as a coupler for selective transfer of an excitation 
between two chains involving an arbitrary odd number of sites $N>3$. 
As shown in Fig. \ref{setup.fig}(b), the coupler involves intermediate sites of the two chains, 
with indices $j_{\rm c}=(N+1)/2$ and $j_\pm=j_{\rm c}\pm 1$.  
The corresponding coupling constants are 
$G_{j_-,j_{\rm c}}^{i,i}=G_{j_{\rm c},j_+}^{i,i}=g$, 
$G_{j_-,j_{\rm c}}^{\rm{s},\rm{d}}=-G_{j_{-},j_{\rm c}}^{{\rm d},{\rm s}}=\kappa$, 
$G_{j_{\rm c},j_+}^{{\rm s},{\rm d}}=-G_{j_{\rm c},j_+}^{{\rm d},{\rm s}}=\kappa$.
In spin networks, the adjustment of geometric phases is possible by looping around magnetic 
fields along the relevant sections \cite{KayEr05}, while for optical networks one may use 
phase shifters.

\begin{figure}
\includegraphics[width=8.cm]{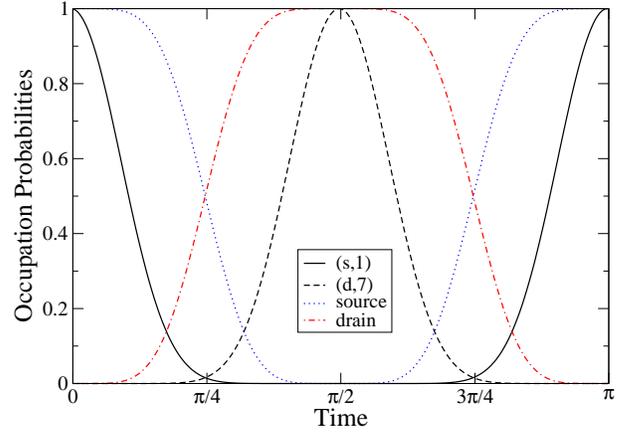}
\caption{
(color online). Time evolution of an excitation in the QDC of Fig. \ref{setup.fig}(b), 
with $N=7$.  The excitation is transferred from the input port $({\rm s},1)$, to the output port of the drain 
channel $({\rm d},N)$ at $t=\tau$. The time is in units of $\Omega^{-1}$. 
} \label{plot1.fig}
\end{figure}

Outside the coupler, only neighboring 
sites of the same channel are coupled according to Eq. (\ref{Omeg}), 
i.e., we have $G_{j,j+1}^{i,i}=\Omega\sqrt{j(N-j)}$ for $i=\{{\rm s}, {\rm d}\}$. 
A qubit state initially prepared at the input port $({\rm s},1)$, 
can be transferred selectively to either of the two output ports 
$\{({\rm s},N),({\rm d},N)\}$ at time $t=\tau$, by adjusting the ratio $g/\kappa$. 
In particular, the transformation (\ref{transAM1}) is performed for $g=\Omega\sqrt{(N^2-1)/4}$ 
and $\kappa=0$, whereas transformation (\ref{transAM2}) takes place for 
$g=\kappa=\Omega\sqrt{(N^2-1)/8}$. 
For the sake of illustration, in Fig. \ref{plot1.fig} we present numerical results pertaining 
to the transfer of a single excitation between two chains of $N=7$ sites each. 
The results have been obtained through the solution of the Schr\"odinger 
equation in the computational basis. The excitation, which occupies initially the site $({\rm s},1)$, 
splits into two parts at the entrance of the coupler (not shown here). 
The two parts follow different paths 
and they split into smaller fractions in the middle of the coupler. 
The various fractions acquire different phases as they propagate through the sections 
of the coupler, and they interfere constructively only on the site $({\rm d}, j_+)$.
In closing, it is worth pointing out that a four-site configuration 
similar to Fig. \ref{setup.fig}(a) may operate as a Hadamard gate \cite{KayEr05}, and 
one may consider judicious combinations of such gates for directional coupling between two channels  
as well. Our three-site configuration, however, cannot be expressed in terms of Hadamard gates.

{\em Summary and outlook.}--- 
We have introduced the notion of dual-channel QDC 
in the context of quantum computation and communication, presenting 
also a general mathematical analogy to a composite spin system.
Employing this analogy, we have been able to 
specify criteria for perfect and deterministic directional coupling of ``flying''-qubit states 
\cite{Vinc00}, between two quantum channels that rely on existing schemes for state transfer.
The present work does not cover all the possible solutions to the problem of 
directional coupling, which is very general and is not associated with a 
particular coupling configuration. 
A number of interesting questions such as the existence of other configurations 
for directional coupling between two or more quantum channels, 
the transfer of arbitrary multiqubit states, the effect of imperfections,  
as well as the extension of (de)multiplexing processes to the quantum world, 
deserve further investigation. 

The work was supported in part by the EC RTN EMALI. I am grateful to P. Lambropoulos 
for helpful comments and discussions.

\end{document}